%% file: wiener_two_messenger.tex
\documentclass[usenatbib]{mnras}

\usepackage{natbib}
\usepackage{aas_macros}
\usepackage{tikz}

\title[Preconditioner-free Wiener filtering with a dense noise matrix]{Preconditioner-free Wiener filtering with a dense noise matrix}

\author[K.~M.~Huffenberger]{Kevin M. Huffenberger
  \\
  Florida State University}

\begin{document}

\maketitle

\begin{abstract}
   This work extends the \citet{2013A&A...549A.111E} iterative method for efficient, preconditioner-free Wiener filtering to cases in which the noise covariance matrix is dense, but can be decomposed into a sum whose parts are sparse in convenient bases.
  The new method, which uses multiple messenger fields, reproduces Wiener filter solutions for test problems, and we apply it to a case beyond the reach of the \citet{2013A&A...549A.111E} method.
  We compute the Wiener filter solution for a simulated Cosmic Microwave Background map that contains spatially-varying, uncorrelated noise, isotropic $1/f$ noise, and large-scale horizontal stripes (like those caused by the atmospheric noise).  We discuss simple extensions that can filter contaminated modes or inverse-noise filter the data.
These techniques help to address complications in the noise properties of maps from current and future generations of ground-based Microwave Background experiments, like Advanced ACTPol, Simons Observatory, and CMB-S4.
\end{abstract}

\begin{keywords}
methods: data analysis -- methods: statistical -- cosmic background radiation 
\end{keywords}

\section{Introduction}

The Wiener filter \citep{wiener1949} is a general method to estimate a stochastic field based on noisy or incomplete data.  It has long been used in astrophysics and cosmology to reconstruct sparse data  \citep[e.g.][]{1992ApJ...398..169R}, denoise maps of the Cosmic Microwave Background \citep[CMB, e.g.][]{1994ApJ...432L..75B} and large scale structure \citep[e.g.][]{1995ApJ...449..446Z}, estimate power spectra  \citep[e.g.][]{1999ApJ...510..551O}, and combine lensing and direct probes of the large scale structure \citep[e.g.][]{2013A&A...560A..33S,2016MNRAS.455.4452A}, among other uses.  It closely relates to Bayesian methods for map and power spectrum estimation \citep[e.g. Gibbs sampling,][]{2004PhRvD..70h3511W,2004ApJS..155..227E}.

The Wiener filter examines a data vector ($d$) that is a linear combination of signal ($s_{\rm true}$) and stochastic noise ($n$):
\begin{equation}
  d = s_{\rm true}+n.
\end{equation}
If the signal covariance is $\mathbf{S}$ and the noise covariance is $\mathbf{N}$, the Wiener filter estimate for the signal is  
\begin{equation}
   s =  \mathbf{ S}(\mathbf{S} + \mathbf{N})^{-1} d,
  \label{eqn:wiener-alt}
\end{equation}
or equivalently
\begin{equation}
   s = (\mathbf{S}^{-1} + \mathbf{N}^{-1})^{-1} \mathbf{ N}^{-1} d.
  \label{eqn:wiener}
\end{equation}
For a particular realization of the data, the Wiener filter solution minimizes:
\begin{equation}
  \chi^2(s) = s^\dag  \mathbf{S}^{-1} s + (d - s)^\dag  \mathbf{ N}^{-1} (d - s),
\end{equation}
maximizing the multivariate Gaussian probability $\propto \exp(-\chi^2/2)$.
No other linear solution has reconstruction errors $\epsilon = s - s_{\rm true}$ with smaller mean-square deviations: the Wiener filter minimizes $\langle \epsilon^\dag\epsilon \rangle$, averaged over all signal and noise realizations.

For large data sets, the solution is difficult unless both $\mathbf{S}$ and $\mathbf{N}$ matrices are sparse (making them easier to invert) and sparse in the same basis (making the sums in equations~\ref{eqn:wiener-alt} and \ref{eqn:wiener} easier to tabulate and invert).  In cosmological contexts, the signal covariance is often isotropic, and so is diagonal in harmonic space.  The noise covariance, by contrast, is often linked to the survey strategy, and so may be dominantly diagonal in real space.  This mismatch has meant that Wiener filter solutions have  traditionally required sophisticated linear system solvers that avoid the inversions of dense matrices, using for example iterative conjugate gradient methods  \citep{2004PhRvD..70j3501H,2007PhRvD..76d3510S}, and these require careful attention to their preconditioners.

\citet{2013A&A...549A.111E} circumvented this difficulty by devising an iterative method to solve the Wiener filter equation without a preconditioner in cases where the signal and noise are sparse in different bases.  To do so they introduce an auxiliary \emph{messenger field}, which splits off a homogeneous component of the noise covariance.  Such a component is diagonal in every orthogonal basis.   This general formalism provides new tools to tackle problems with isotropic signal and spatially varying but uncorrelated noise, and also closely relates to a method for Gibbs sampling.   Already various authors have employed these methods in the context of CMB polarization \citep{2013A&A...549A.111E}, CMB gravitational lensing \citep{2015ApJ...808..152A}, large-scale structure \citep{2015MNRAS.447.1204J,2016MNRAS.455.3169L}, and cosmic shear \citep{2016MNRAS.455.4452A}.  In all of these cases the signal covariance is (block-) diagonal in harmonic space and the noise covariance is (block-) diagonal in pixel space. These methods are under rapid development, and recently \citet{2017arXiv170208852K} showed that additionally splitting off a homogeneous portion of the signal covariance can help to speed the convergence. 

In this work we extend this type of solution to a new class of problems.  We consider cases in which the noise covariance is dense, but is a sum of pieces that can each be represented sparsely in convenient bases.  Our approach achieves this by adding messenger fields for each additional term required for the covariance.

This paper is organized so that in section \ref{sec:methods} we review the  \citet{2013A&A...549A.111E} method and introduce our extension.  In section \ref{sec:results} we apply our method to simulations with progressively more complex noise.  In section \ref{sec:discussion} we discuss the implications. An appendix provides more details of our approach.  

\section{Methods}\label{sec:methods}

\citet{2013A&A...549A.111E} (hereafter EW) showed that the Wiener filter equation could be recast as a system of equations with the same solution.  For a matrix $\mathbf{T}$ (which is arbitrary except that $\mathbf{\bar N = N - T}$ should be positive definite), the solution for the Wiener filter also solves
\begin{eqnarray}
   t &=& (\bar  \mathbf{N}^{-1} +  \mathbf{T}^{-1})^{-1} \left( \mathbf{ \bar N}^{-1} d + \mathbf{ T}^{-1} s \right) \label{eqn:ew_t}\\
    s &=& (\mathbf{S}^{-1} +  \mathbf{T}^{-1})^{-1}\mathbf{ T}^{-1} t \label{eqn:ew_s}
\end{eqnarray}
for a unique $t$. (The value of $t$ depends on our choice for $\mathbf{T}$.)  Furthermore, iterating  equations (\ref{eqn:ew_t})--(\ref{eqn:ew_s}) will exponentially converge to the Wiener solution.  Here we solve for two vectors ($s,t$) instead of one, which makes the problem look harder until we notice that we can make specific, convenient choices for $\mathbf{T}$.  If $\mathbf{T}$ is proportional to the identity matrix ($\mathbf{T} = \tau_{\rm EW} \mathbf{I}$, where $\tau_{\rm EW}$ is a scalar constant) then it is sparse in every orthonormal basis.  Thus if $\mathbf{S}$ and $\mathbf{N}$ can separately be written in distinct sparse bases, a combination like $(\mathbf{S}^{-1} + \mathbf{T}^{-1})$ can be written in a sparse way and easily inverted. That allows the equations to be evaluated directly without recourse to linear system solvers.  This is especially quick if the signal and data vectors can be transformed to the convenient bases via a fast transform (e.g. between real and harmonic space).  A ``cooling schedule'' that artificially scales up the $\mathbf{T}$ covariance at the start, then gradually returns it to the proper value, can help speed the convergence.  

Although the EW equations (\ref{eqn:ew_t})--(\ref{eqn:ew_s}) are always valid, it only makes sense to apply them in cases when both the signal and noise have convenient sparse representations that allow the above shortcut.  However, for many problems of interest, the noise covariance is dense for all convenient bases.  One example problem is the optimal filtering of maps from ground-based CMB experiments, where a significant amount of the noise comes from the atmosphere.  That portion of the noise covariance is strongly correlated in real space between different parts of the map, while the portion of the noise covariance due to detector noise and foreground masking are sparse only in real space.  Thus the noise covariance is dense both in real and harmonic space and is by itself difficult to invert.

Here we show that we can make progress in this case if we can decompose the noise covariance into a sum of simpler terms.  If each term in the sum is sparse in some---possibly distinct---basis, we can write down an EW-style set of coupled algebraic equations that use messenger fields to iteratively produce the Wiener filter solution.  This requires us to add additional messenger fields.  In the case where the covariance is broken into two pieces, $\mathbf{N} = \mathbf{N}_0 + \mathbf{N}_1 $, we can define two messenger fields ($t_0, t_1$ with covariances $\mathbf{T_0,T_1}$) and an auxiliary (data-like) field $d_0$.  Then the following set of equations iteratively yields the solution to the Wiener filter equation, in analogy to the EW method:
\begin{eqnarray}
   t_1 &=&(\bar  \mathbf{N}_1^{-1} +  \mathbf{T}_1^{-1})^{-1} \left( \bar  \mathbf{N}_1^{-1} d + \mathbf{ T}_1^{-1} d_0 \right)  \label{eqn:2mess_t1}  \\
   d_0 &=& (\bar  \mathbf{N}_0^{-1} +  \mathbf{T}_1^{-1})^{-1} \left( \mathbf{ \bar N}_0^{-1} t_0 + \mathbf{ T}_1^{-1} t_1 \right) \label{eqn:2mess_d0} \\
   t_0 &=& (\bar  \mathbf{N}_0^{-1} +  \mathbf{T}_0^{-1})^{-1} \left( \mathbf{ \bar N}_0^{-1} d_0 + \mathbf{ T}_0^{-1} s \right) \label{eqn:2mess_t0} \\ 
    s &=& ( \mathbf{S}^{-1} +  \mathbf{T}_0^{-1})^{-1}\mathbf{ T}_0^{-1} t_0. \label{eqn:2mess_s}
\end{eqnarray}
Appendix \ref{sec:derivation} describes the Bayesian hierarchical model that is the origin of these equations.
Each additional piece of the noise covariance will describe two additional ($d_0$- and $t$-like) fields and equations that we can chain together to get from the data to the signal estimate.  Choosing $\mathbf{T}_0 = \tau_0 \mathbf{I}$ and $\mathbf{T}_1 = \tau_1 \mathbf{I}$ makes the matrix inversions in equations~(\ref{eqn:2mess_t1})--(\ref{eqn:2mess_s}) sparse and trivial in the proper basis, and allows iterative solutions to a much broader class of problems.  The convergence proceeds fastest when these $\tau_0,\tau_1$ parameters are as large as possible, and match the minimum eigenvalues of $\mathbf{N}_0$ and $\mathbf{N}_1$ respectively. (Any larger and the $\bar \mathbf{N}$ matrices are not positive definite.)  In some cases these equations can be combined to speed the convergence (appendix \ref{sec:derivation}).

In our implementations, we use equations~(\ref{eqn:2mess_t1})--(\ref{eqn:2mess_s}) directly and initialize $s = t_0 = d_0 = 0$, but this is not critical to converge to the solution.  For the problems of interest in this paper, we let $\mathbf{N}_1$ represent a real-space component of the noise on the full sky and  $\mathbf{N}_0$ represent a spherical-harmonic-space component.  The signal covariance is isotropic and sparse in harmonic space for all the cases we consider.  We compute equation~(\ref{eqn:2mess_t1}) in pixel space and equations~(\ref{eqn:2mess_d0})--(\ref{eqn:2mess_s}) in harmonic space, where the matrices are diagonal and trivially invertible.    Because the Wiener-filtered solution is ultimately computed in harmonic space in this implementation, it has a strict band limit.

The main computational cost is in the spherical harmonic transforms.  Two transforms are required per iteration in the test problems that follow, one to take $t_1$ to harmonic space and one to take $d_0$ to real space.  This is the same number of transforms required by the EW method for similar problems on the sky.

When we implement a cooling scheme, we set $\tau_0 \rightarrow \lambda_0 \tau_0$ and  $\tau_1 \rightarrow \lambda_1 \tau_1$.  The cooling parameters $\lambda_0,\lambda_1$ start large so that the $\mathbf T$ matrices dominate the other covariances (the $\bar \mathbf{N}$ matrices are left fixed), and are gradually lowered to unity to achieve the Wiener filter solution.

We want to monitor the quality of the solution as it is converging to help us set a schedule for the cooling parameters. An obvious candidate for this job is $\chi^2(s)$, but unfortunately for this case, it is difficult to compute the second term in
\begin{equation}
  \chi^2(s) = s^\dag \mathbf{S}^{-1} s + (d-s)^\dag(\mathbf{N}_0 + \mathbf{N}_1)^{-1}(d-s),
\end{equation}
because we do not have a quick way to invert the sum of the noise matrices, as they are sparse in different bases.  Conjugate gradient descent would work to compute $\chi^2$, but is not practical as a way to monitor the progress of our iterative solution, which purposefully avoids gradient descent techniques.

To help, we invert equation (\ref{eqn:wiener-alt}) to introduce a reconstructed data field, which is a function of our Wiener-filter solution and is easy to compute:
\begin{equation}
  d_{\rm rec}(s) = (\mathbf{S} + \mathbf{N}) \mathbf{S}^{-1} s =  (\mathbf{I} + \mathbf{N} \mathbf{S}^{-1}) s.
\end{equation}
In practice, we treat the harmonic-space signal and noise covariances with a band limit, so not all parts of the data are recoverable, and we do not expect to reconstruct the data exactly, particularly on small scales.  Even so, the difference between the reconstructed data and the actual data,
\begin{equation}
\Delta d = d_{\rm rec} - d,
\end{equation}
is a useful and practical metric that allows us to set the cooling schedule.

\section{Results} \label{sec:results}

We consider three test problems to probe the new method for Wiener filtering.  This first is trivial, containing uniform white noise, and can be solved directly as well as by the iterative methods.  The second has non-uniform, uncorrelated noise, which both the EW method and the new method can handle.  The third addresses an application with a dense, composite noise matrix that is suitable for our extension but cannot be handled by the original EW method.

\subsection{Tests with homogeneous, uncorrelated noise}
In the trivial case of homogeneous noise, we can directly solve the Wiener filter (equation~\ref{eqn:wiener}) in harmonic space, so it makes a sensible starting point to verify our Wiener filter solution.  We constructed a first test problem on the sphere so that the data on large scales was dominated by an isotropic signal (with a red power spectrum), while on small scales it was dominated by uniform white noise, added in real space with covariance $\mathbf{N} = (\sigma^2/\Omega_{\rm pix}) \mathbf{I}$.  (Equivalently, the noise has power spectrum $N_l = \sigma^2$.)  Then we solved for the Wiener-filtered map three ways: (1) directly in harmonic space, (2) using the EW iterative method, and (3) using our multiple messenger field method.  We used a HEALPix\footnote{\url{http://healpix.sourceforge.net}} pixelization at $N_{\rm side} = 64$ resolution.  All harmonic space computations were limited to $l \leq 3 N_{\rm side} = 192$.

Both iterative methods converge to the direct, harmonic space solution.  For the EW method, in the limit $\tau_{\rm EW} \rightarrow \sigma^2$, the messenger field represents all the uniform noise, and the method reduces to the direct solution and trivially converges in a single step.  For this problem the $\chi^2$ for the EW method is slightly better than for the harmonic space solution ($\Delta \chi^2 / \chi^2 \sim \mbox{few} \times 10^{-6}$), perhaps because the EW solution partially handles the noise in pixel space where the noise is generated, and the harmonic space solution includes only noise power up to the band limit.

The implementation of the multiple-messenger method converges to the same solution, but the uniform noise case is awkward for it because we need to make a decision about how to split the noise power between the $\mathbf{N}_0$ and $\mathbf{N}_1$ parts of the covariance. (The code fails if one of the covariances is left completely empty.)  Putting the bulk of the white noise into $\mathbf{N}_0$ gives the fastest convergence: if a ninety-nine percent of the noise power goes into $\mathbf{N}_0$ the $\chi^2$ value converges to a part in $10^3$ after four iterations, but if only half the power is assigned to $\mathbf{N}_0$, this level of convergence takes 75 iterations without any cooling scheme.  This method converges to a $\chi^2$ better than the harmonic solution, but not as good as the EW solution (and again, $\chi^2$ values for all solutions have only slight differences, $\Delta \chi^2 / \chi^2  \sim \mbox{few} \times 10^{-6}$).

\subsection{Tests with inhomogeneous, uncorrelated noise}

Our second test is the case that the EW solution handles most straightforwardly: inhomogeneous but uncorrelated noise.  We modify the above test data set to include a mask (where the noise covariance for pixels is infinite, and the inverse covariance is zero) and to boost the noise in pixels near the mask.  Both the EW method and the multiple messenger field method converge to the same solution.  Empirically the convergence of the new method's $\chi^2$ is exponential (as it is for the EW method).  The rate of convergence depends again on the decomposition of the noise matrix and was about about 100 times slower than EW in some cases, although we did not try to optimize for speed or implement any cooling for this comparison.  The noisy regions at the mask edge were the slowest to converge.

\subsection{Application with dense noise}

\begin{figure}
  \includegraphics[width=\columnwidth]{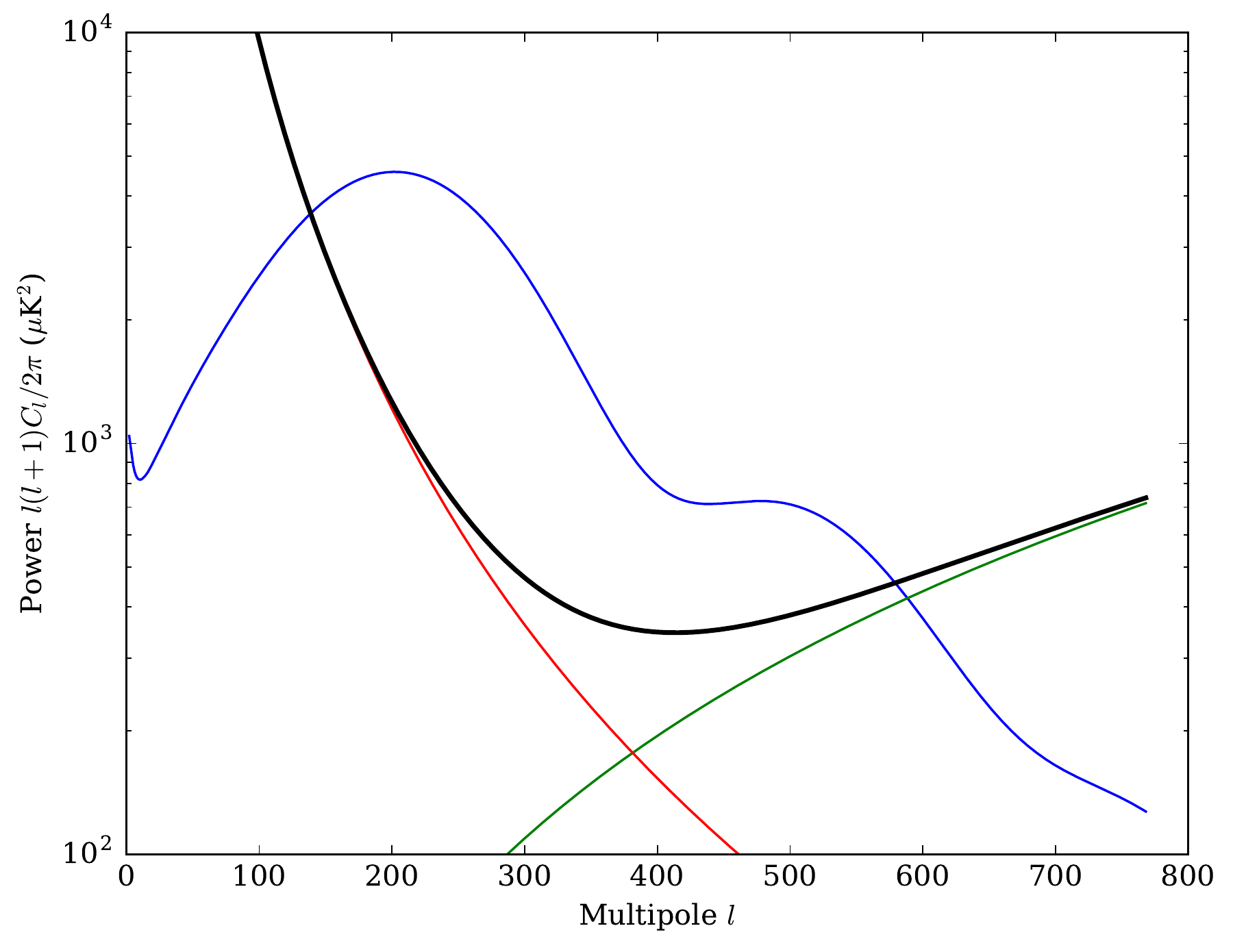}
  \caption{Input power spectra of signal and noise for our example with dense noise.  The blue line is the signal, a CMB temperature power spectrum in a standard $\Lambda$CDM cosmology, smoothed with a 0.3 deg full-width-half-maximum beam.  The red line represents power in correlated noise, a fraction of which is allocated to all modes, and the remainder to the $m=0$ modes alone, so that the noise is anisotropic.  The green line represents the white noise level in the cleanest portions of the map.  The black line shows the sum of the noise power for the cleanest portions of the map, but the white noise level is higher in regions near to the edges of the mask.}
  \label{fig:signal-noise}
\end{figure}

Our third test case is one that our new multiple messenger field method can handle, but the EW method cannot.  We construct a noise covariance that is dense both in real and harmonic space.  In this way it mimics ground-based CMB experiments that observe the statistically isotropic microwave background, but also noise that is spatially anisotropic and contains correlations from the atmosphere and scan pattern.  In recent measurements from, for example, ACTPol \citep{2016arXiv161002360L}, atmospheric noise dominates the temperature signal at large scales, while detector noise dominates the signal at small angular scales.  At the edges of the map, where less integration time is spent, the noise level is also higher.  Our test problem tries to incorporate all these features (and a mask) but treats them at reduced resolution (compared to ACTPol) to shorten runtimes for study.  We use  $N_{\rm side} = 256$ resolution and harmonic band limit $l \leq 768$.

\begin{figure*}
  {
    \newlength{\mapwidth}
    \setlength{\mapwidth}{0.495\textwidth}
  \includegraphics[width=\mapwidth]{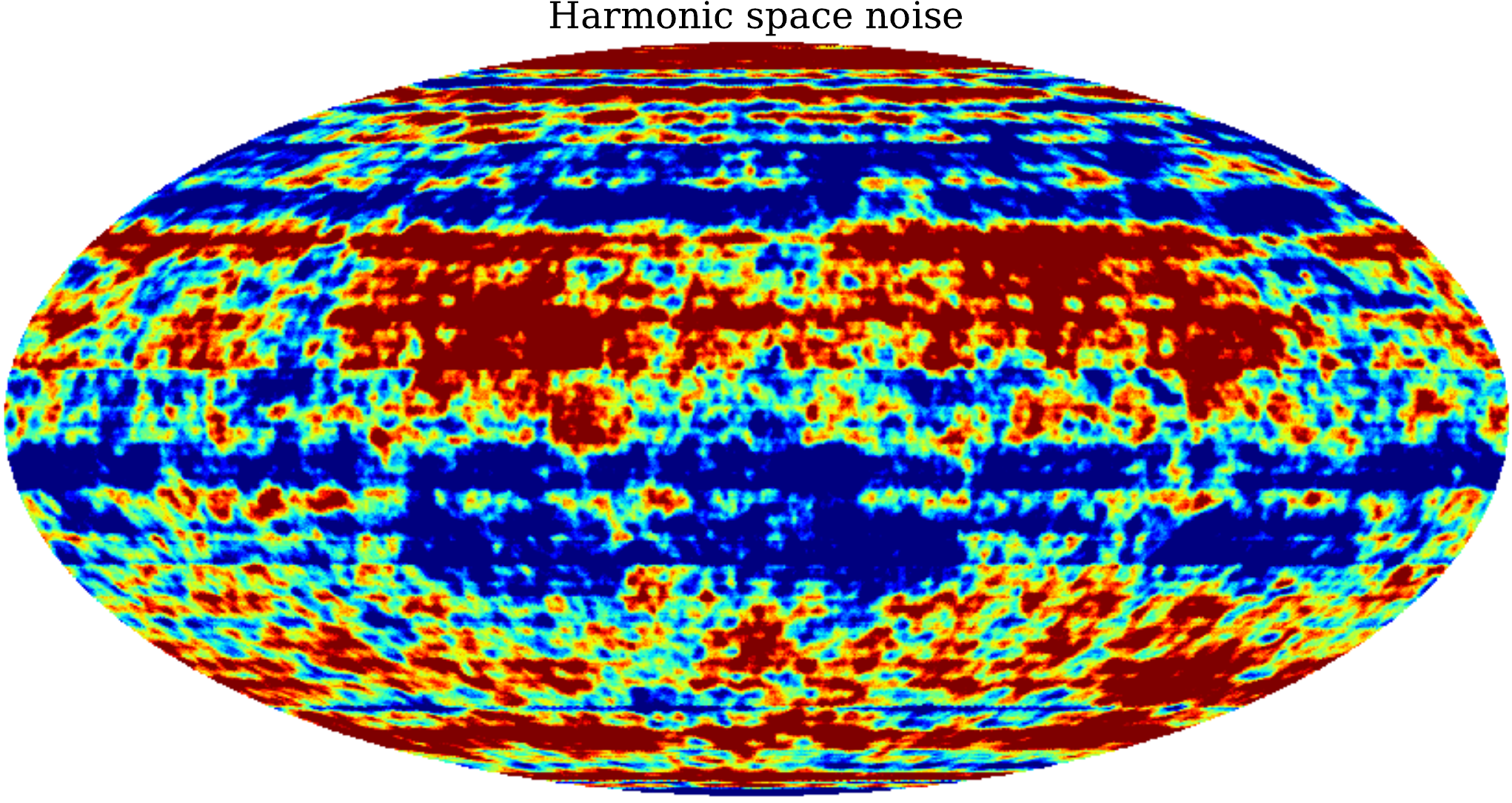}\hfill
  \includegraphics[width=\mapwidth]{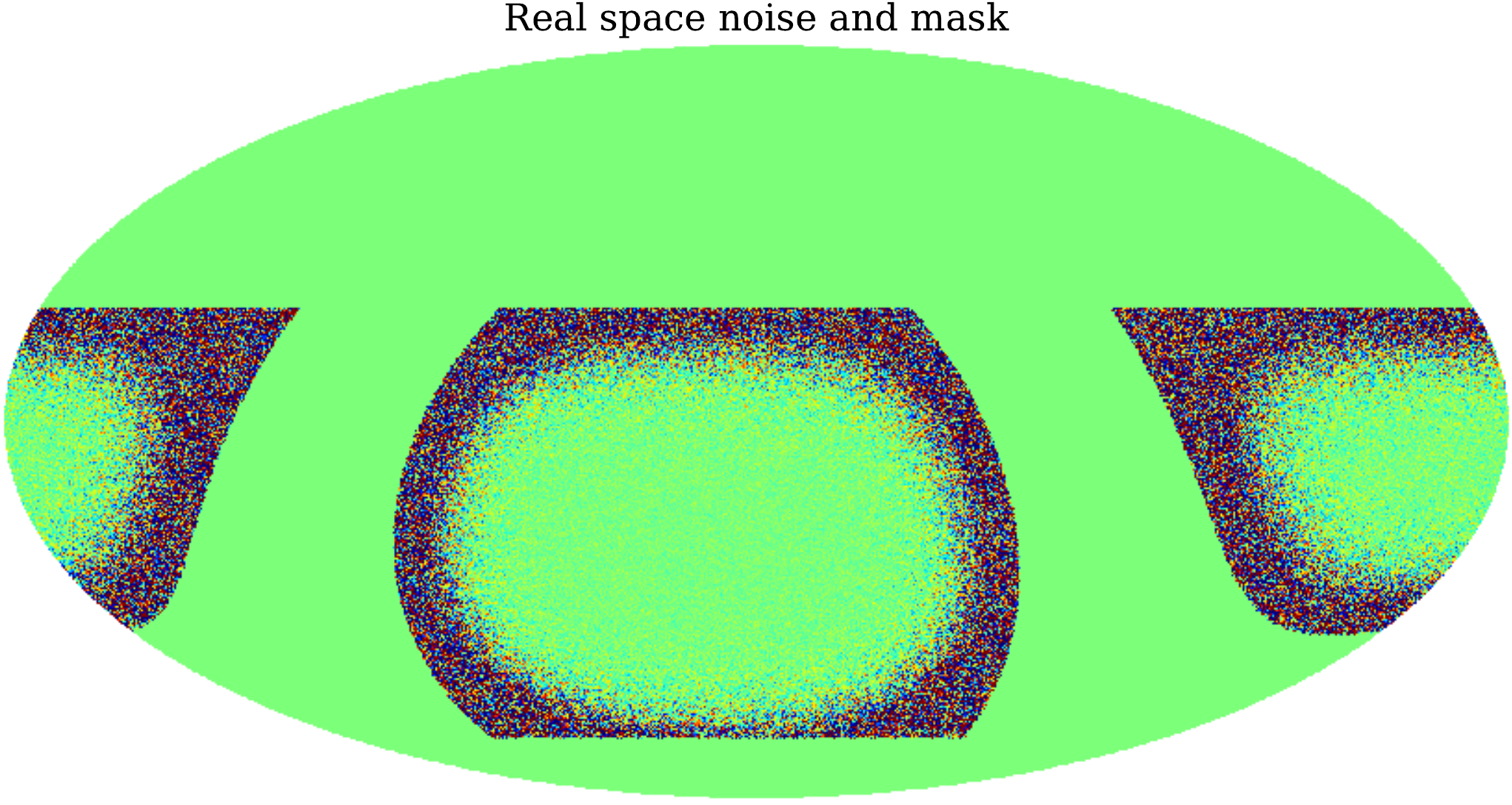}
  \vspace{1ex}
  \includegraphics[width=\mapwidth]{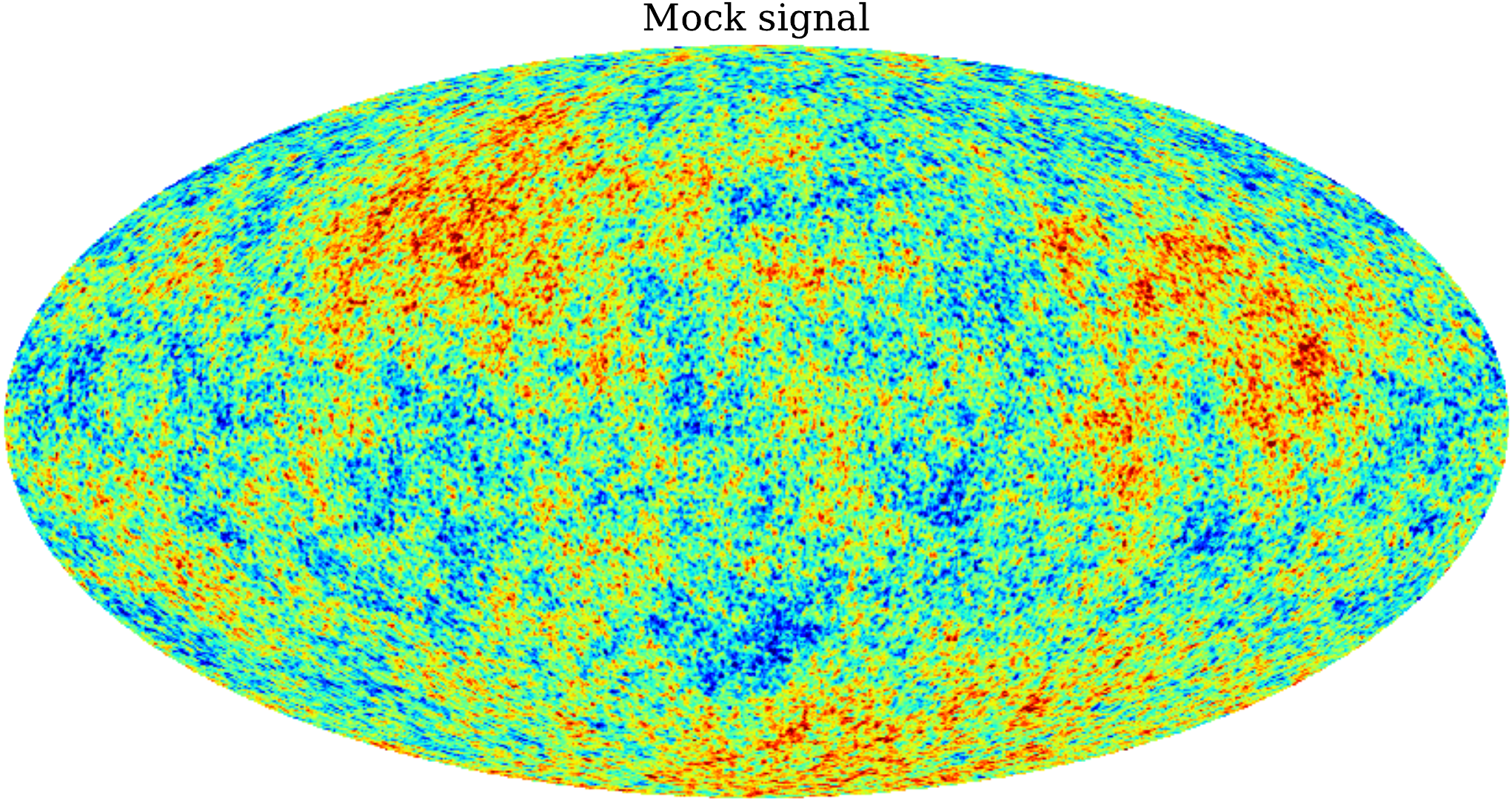}\hfill
  \includegraphics[width=\mapwidth]{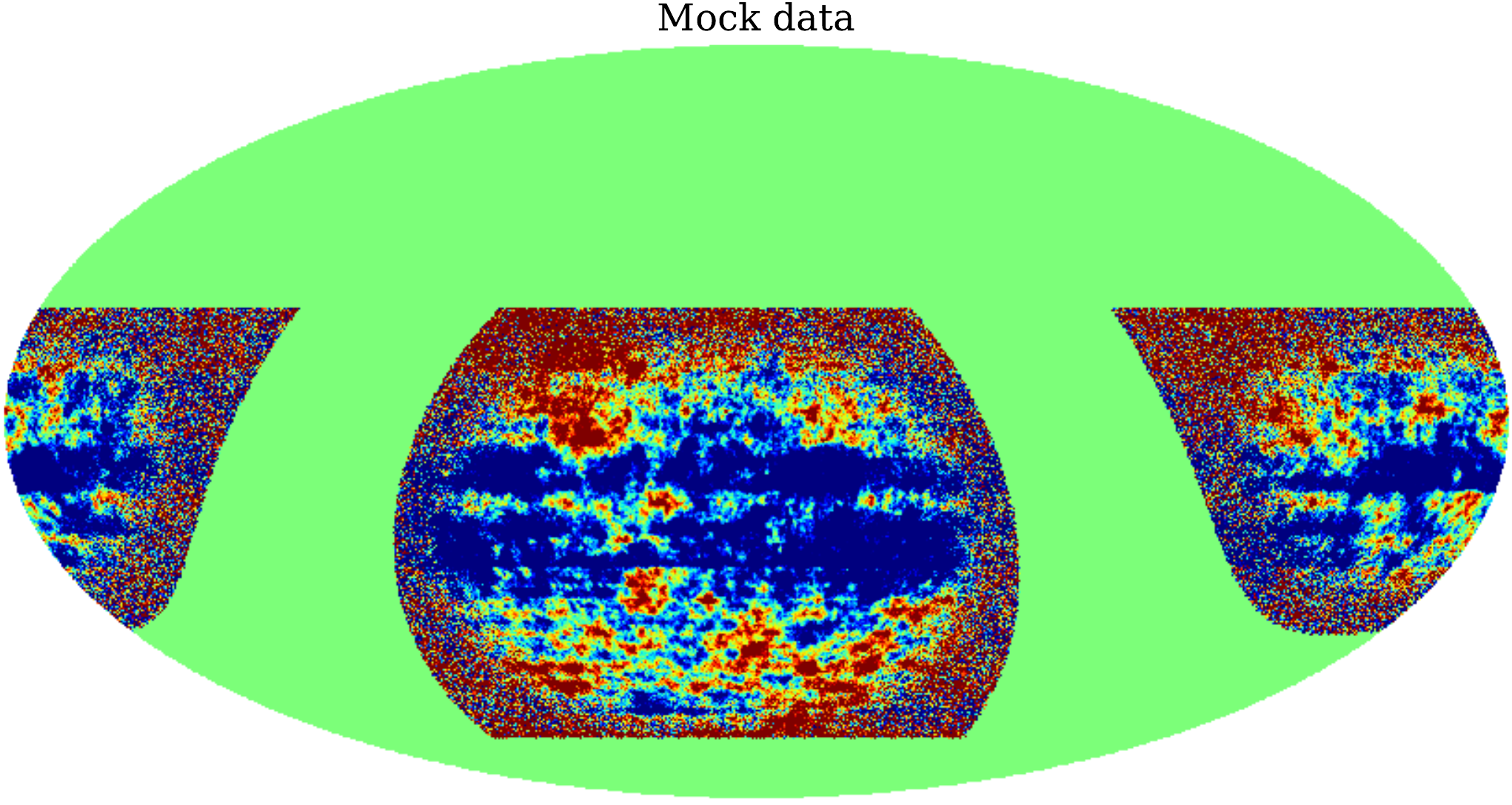}
  \vspace{1ex}
  \includegraphics[width=\mapwidth]{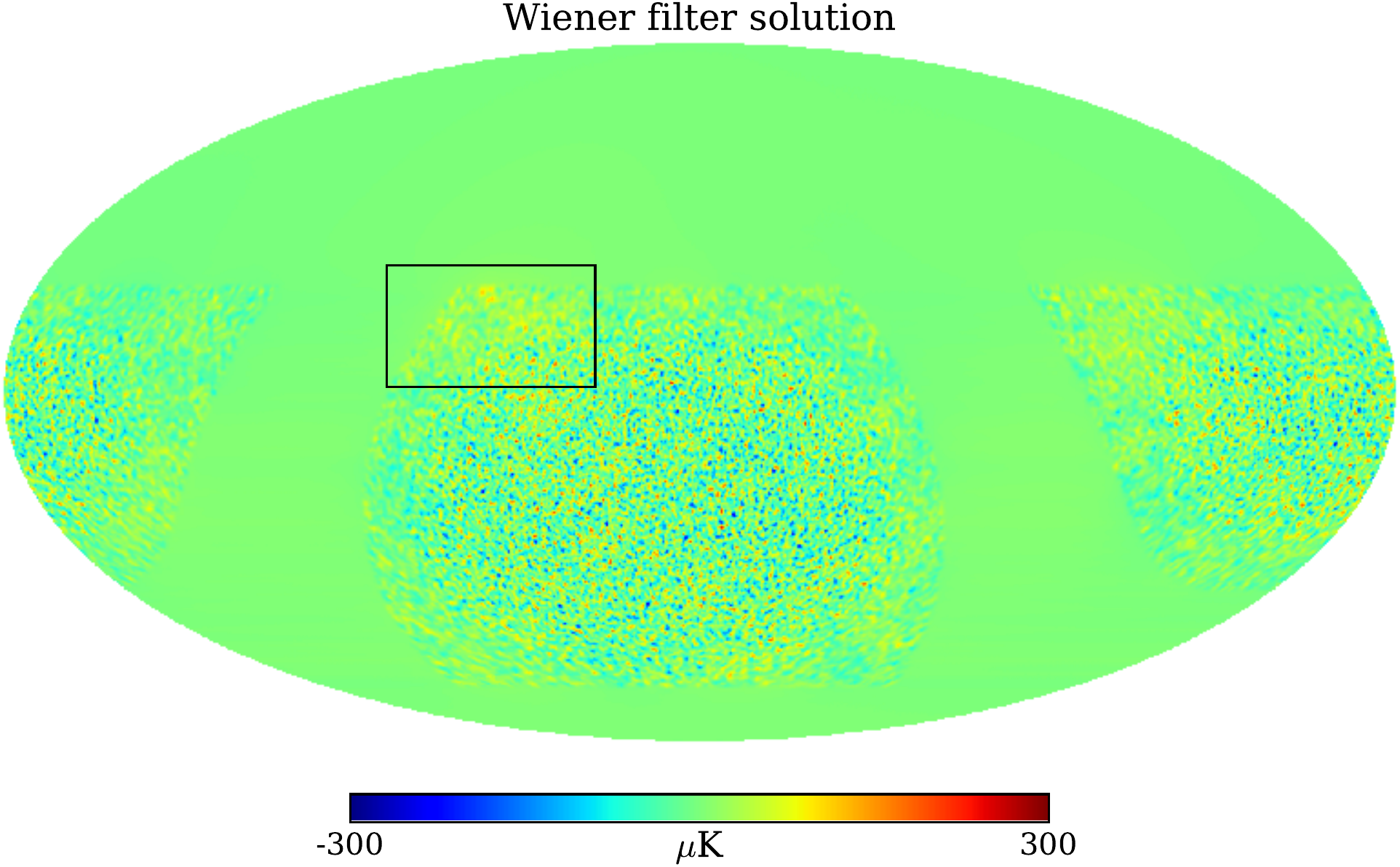} \hfill
  \raisebox{0.6cm}{\includegraphics[width=\mapwidth]{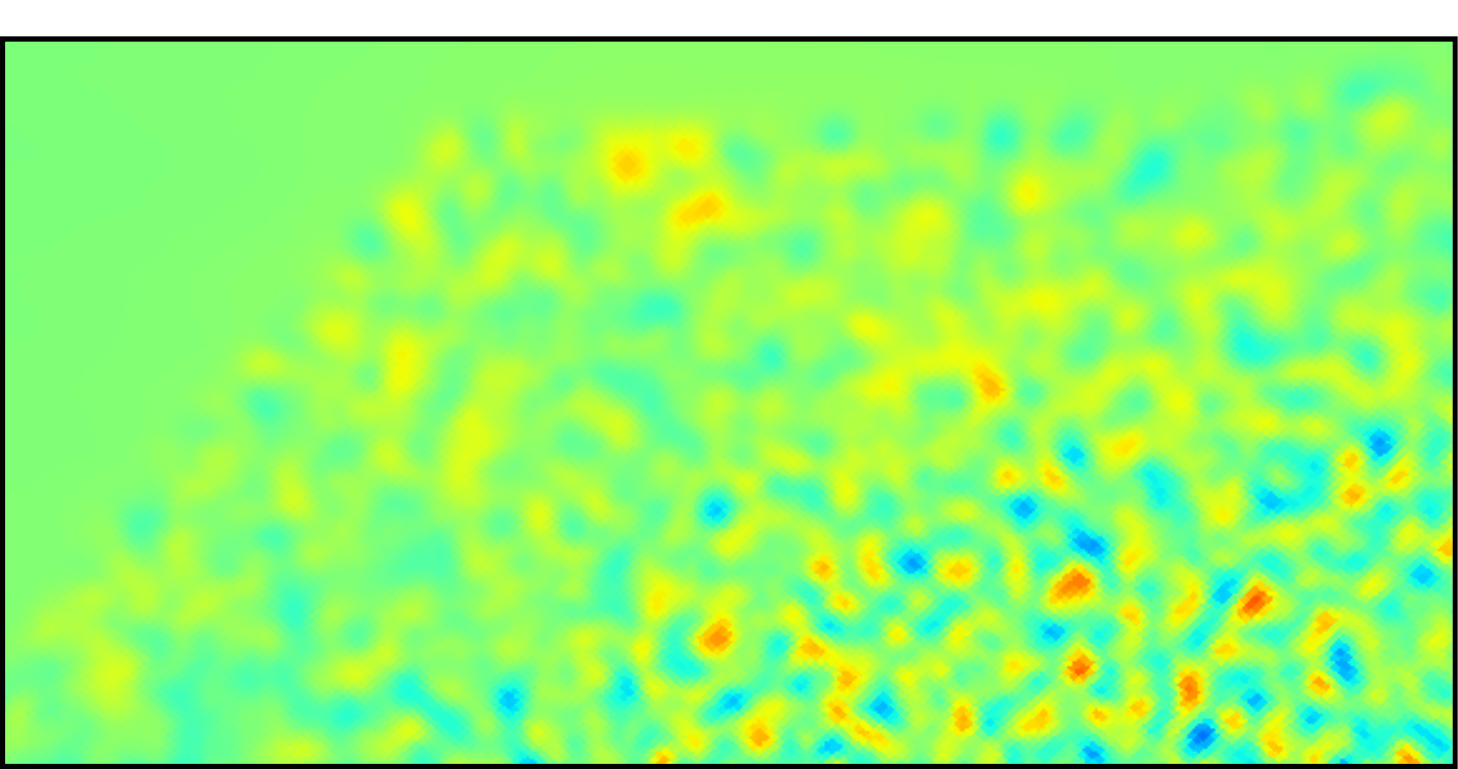}}
  }
  \caption{Maps of noise, signal, data, and the Wiener-filter solution, all in celestial coordinates and on a common color scale.  \textit{Top left and right:} contributions to the map noise generated in harmonic space and in real space. \textit{Middle left:} mock CMB temperature signal. \textit{Middle right:} mock data, which is the sum of both noise contributions and the signal.  \textit{Bottom:} Wiener filter solution.  The box at the right zooms in on the indicated area.}
  \label{fig:maps}
\end{figure*}

We build our mock data from an isotropic signal, a Gaussian random field  ($s_{\rm CMB}$) generated from a $\Lambda$CDM power spectrum for CMB temperature and smoothed with a 0.3 deg full-width-half-maximum beam, plus two noise terms:
\begin{equation}
  d = s_{\rm CMB} + n_0 + n_1.
\end{equation}

The $n_0$ portion of the noise is a Gaussian random field, and has a covariance that is diagonal in harmonic space, but not isotropic:
\begin{equation}
  \langle n_{0,lm} n_{0,l'm'}^* \rangle = N_{0,lm} \delta_{ll'}\delta_{mm'}. 
\end{equation}
We show the shape of the power spectrum for the correlated noise in Figure \ref{fig:signal-noise}, but half of the power is put into the $m=0$ modes alone, which introduces horizontal noise stripes.  We also give this harmonic component of the noise half of the white noise for the cleanest portions of the map.  Figure \ref{fig:maps} shows a realization of this noise component along with the other fields.

The $n_1$ portion of the noise is also a Gaussian random field, but has a covariance that is diagonal is pixel space.   The mask roughly corresponds to the observable sky from the Atacama, excluding the portions outside declination range $-68^\circ < \delta < 22^\circ$.  It also excludes the Galactic plane within $|b| < 20^\circ$.  The mask is enforced by giving masked pixels infinite variance, or in practice setting the inverse variance $ \bar N_{1,pp}^{-1} = 0$ for masked pixel index $p$. Otherwise, the noise is spatially uncorrelated but the pixel variance rises near the edges of the mask.   In the least noisy portions of the sky, the signal power exceeds the noise for $139 < l < 588$.

For the specific example we consider, we start the cooling parameters at $\lambda \sim 5\times 10^5$, and iterate until the quantity $(\Delta d^\dag \Delta d)$ changes by  a fraction smaller than $10^{-2}$ per iteration.  At that time, we reset $\lambda \rightarrow \lambda^{0.7}$, and continue to iterate.  The procedure gradually lowers $\lambda$ to unity.

\begin{figure}
  \includegraphics[width=\columnwidth]{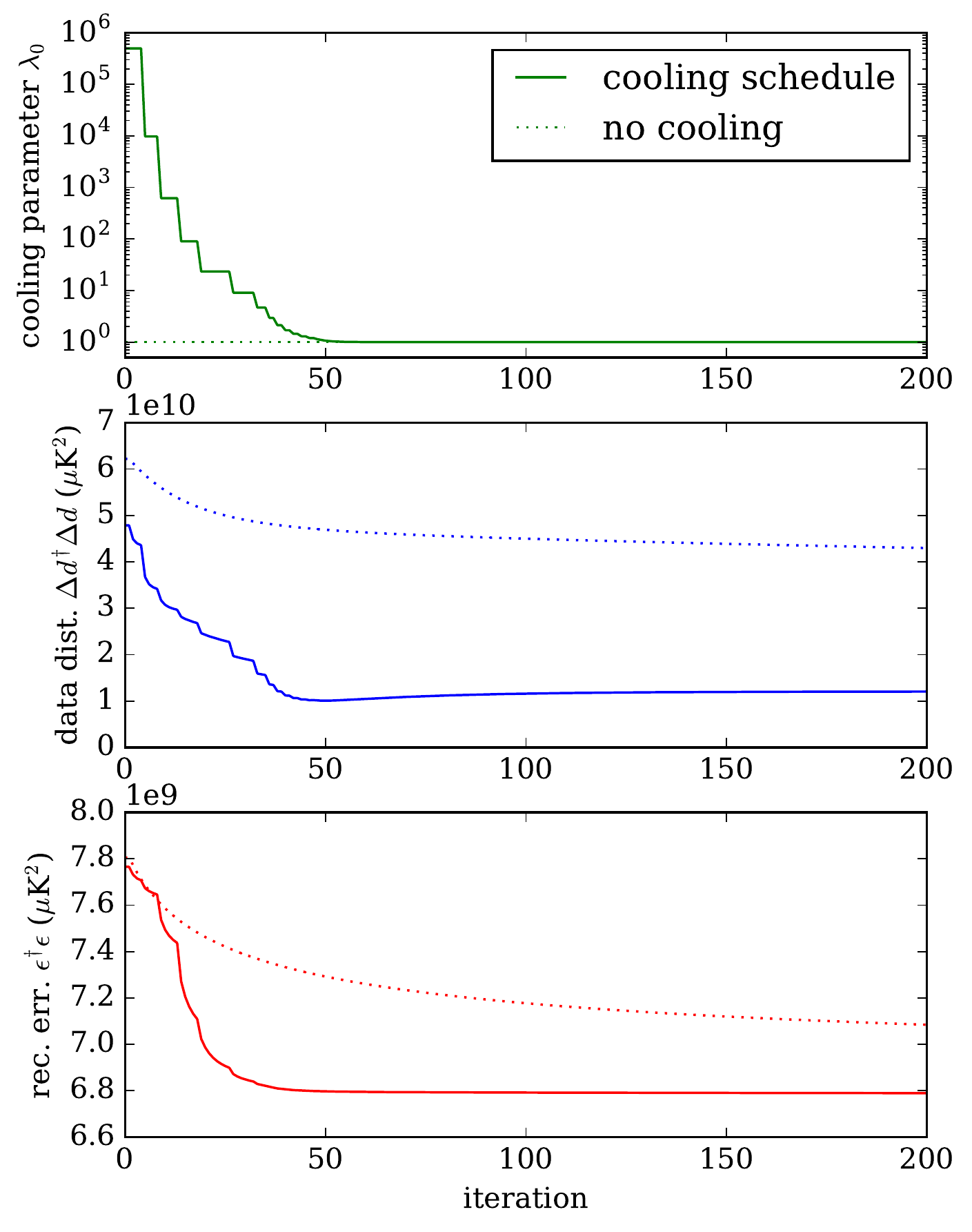}
  \caption{\textit{Top:} One of the cooling parameters that aids the convergence to the Wiener filter solution.  It is gradually lowered to unity as we iterate.  \textit{Middle:} The distance between the reconstructed data and the actual data.  We monitor this quantity, and when it begins to plateau, we lower the cooling parameter. \textit{Bottom:} The reconstruction error of the Wiener-filtered solution compared to the true signal.  Implementation of a cooling scheme is vital to achieve rapid convergence and a quality solution.}
  \label{fig:reconstruction_errors}
\end{figure}

In Figure~\ref{fig:maps} we show the Wiener filter solution, and in Figure~\ref{fig:reconstruction_errors}, we show how the cooling, the reconstruction's distance to the data, and the reconstruction error each converge as we iterate.   The Wiener filter solution has power only on intermediate scales, cleanly approaches zero in the masked regions, and shows little to no residual horizontal striping.  With this specific cooling prescription, the $\lambda$ parameters approach unity after about 50 iterations.  (We have not tried hard to optimize the cooling.)  At roughly that time, the reconstructed data makes its closest approach to the actual data.  It moves away upon further iterations, but the reconstruction error $\epsilon = s - s_{\rm true}$, inaccessible for real data but accessible for our test problem, continues to improve. Implementation of the cooling schedule was vital to obtain a good solution.  Convergence without cooling is very much slower (constant $\lambda = 1$ in Figure~\ref{fig:reconstruction_errors}), and stripes persisted even in the masked regions after a few thousand iterations.


\section{Discussion} \label{sec:discussion}

The test problems demonstrate that our extension to the EW method can compute Wiener filtered solutions, and treat cases with complicated noise properties.  We can straightforwardly extend such an approach to CMB polarization.  Ground-based $B$-mode observatories will provide detailed measurements of the covariance structure of the atmosphere in polarization at microwave frequencies, allowing construction of the proper noise covariance models, although this may require significant work.  Other extensions can apply to CMB lensing or large scale structure measurements that have a combination of local and correlated noise effects.  

Because of the close connection of the Wiener filter both to quadratic power spectrum estimation \citep{1999ApJ...510..551O} and Gibbs sampling \citep{2004PhRvD..70h3511W,2004ApJS..155..227E,2015MNRAS.447.1204J}, multiple-messenger methods may allow improvements to power spectrum estimation for ground-based CMB observations.  We will continue to explore these avenues at higher resolution and in more realistic scenarios applied to, for example, the Simons Observatory and CMB-S4.

The usefulness of this approach ultimately depends on our ability to construct noise covariance matrices as a sum $\mathbf{N} = \sum_i \mathbf{N}_i$ so that the inversions of $\bar \mathbf{N}_i = \mathbf{N}_i - \mathbf{T}_i$ are efficient.  Already in this category are all covariances based on pixel-space and harmonic-space masks and observation hit-count maps.  Parallel noise stripes, as considered here, could be placed in any orientation via straightforward rotations of the map.  Small numbers of specific modes, represented as
\begin{equation}
  \mathbf{N}_i = \mathbf{F} \mathbf{\Lambda} \mathbf{F}^\dag,
\end{equation}
for a small matrix $\mathbf{\Lambda}$, may also be treated efficiently. Inversion via the Woodbury formula,
\begin{eqnarray}
  \bar \mathbf{N}_i^{-1} &=& (-\mathbf{T}_i + \mathbf{F} \mathbf{\Lambda} \mathbf{F}^\dag )^{-1}  \\ \nonumber
  &=& -\mathbf{T}_i^{-1} - \mathbf{T}_i^{-1}  \mathbf{F} ( \mathbf{\Lambda}^{-1} -  \mathbf{F}^\dag \mathbf{T}_i^{-1}  \mathbf{F} )^{-1}  \mathbf{F}^\dag  \mathbf{T}_i^{-1}
\end{eqnarray}
is efficient because the term in parenthesis is a small matrix (with the dimension of $\mathbf{\Lambda}$), and the other matrix multiplications and inversions of $\mathbf{T}$ are trivial.

Finally, deliberate modifications of the covariance matrices away from their realistic values can achieve other desirable outcomes.   As is standard practice, intentionally letting the covariance for specific (perhaps untrustworthy) modes go to infinity will project out those modes entirely.  This strategy can downweight large-scale ground pickup or sidelobe features. For another example, we could set the signal covariance to be artificially small and diagonal, say $\mathbf{S} = \alpha \mathbf{I}$ for sufficiently small $\alpha$, to allow computation of approximately inverse noise weighted maps in cases where the noise covariance is not easily inverted.  By casting it as a Wiener filtering problem,
\begin{equation}
 \mathbf{N}^{-1} d \approx  \alpha^{-1} \left[ \alpha \mathbf{I} (\alpha \mathbf{I} + \mathbf{N})^{-1} d \right],
\end{equation}
these multiple-messenger methods can find an approximate solution, given a suitable decomposition of $\mathbf{N}$.

\section*{Acknowledgments}
KMH thanks Aditya Rotti for valuable discussions during the course of this work.
KMH also thanks Benjamin Wandelt for providing useful comments on a draft of this paper.
We acknowledge support from the NASA ATP program under grant NNX17AF87G.
Some of the results in this paper have been derived using the HEALPix package \citep{2005ApJ...622..759G}.

\appendix
\section{Iterative Wiener filtering for a composite noise covariance} \label{sec:derivation}

The Wiener filter maximizes the Gaussian probability of the signal given the data, signal covariance, and noise covariance.  The logarithmic probability in this case, up to a normalization constant, is 
\begin{equation}
-2 \log P(s,d) = s^\dag \mathbf{S}^{-1} s + (d-s)^\dag\mathbf{N}^{-1}(d-s) + \mbox{const.} \label{eqn:wiener_prob}
\end{equation}
We obtain the Wiener filter solution by finding the signal vector which maximizes the conditional probability for fixed data, $P(s|d)$.  We proceed by completing the square on the right hand side, which yields
\begin{equation}
-2 \log P(s|d) = (s - \mu_s) ^\dag (\mathbf{S}^{-1} + \mathbf{N}^{-1}) (s - \mu_s) + \mbox{const.}
\end{equation}
where $\mu_s = (\mathbf{S}^{-1} + \mathbf{N}^{-1})^{-1} \mathbf{ N}^{-1} d$.  Thus the mean (maximum probability) point of the distribution, $s=\mu_s$,  is the Wiener solution in equation~(\ref{eqn:wiener}).

We can arrive at the \citet[][hereafter EW]{2013A&A...549A.111E} solution by first noting that the augmented Gaussian probability distribution
\begin{eqnarray}
  -2 \log P_1(s,d,t) &=& s^\dag \mathbf{S}^{-1} s \\ \nonumber
  &&+ (t-s)^\dag\mathbf{T}^{-1}(t-s)  \\ \nonumber
  &&+ (d-t)^\dag \mathbf{\bar N} ^{-1}(d-t)+ \mbox{const.}
\end{eqnarray}
is equivalent to the above equation~(\ref{eqn:wiener_prob}) for $\mathbf{\bar N = N -T}$ after the messenger field $t$ is marginalized out.  In other words,
\begin{equation}
\int dt\, P_1(s,d,t) = P(s,d).
\end{equation}
The distribution is Gaussian, and so marginalizing one variable does not change the maximum probability positions for the other variables.  Thus the value of $s$ that maximizes the probability is the same in both distributions $P_1$ and $P$.
 
Completing the square for $t$ and $s$ respectively yields the EW solution (our equations~(\ref{eqn:ew_t})--(\ref{eqn:ew_s})).  By iterating, we repeatedly find the maximum probability point along slices of constant $s$ and $t$, and step our way to the maximum probability for the joint distribution $P_1$, which in the end gives the same value for the signal $s$ as the Wiener solution from $P$.

Furthermore, \citet{2015MNRAS.447.1204J} showed that in a Bayesian hierarchical framework, this scheme trivially lends itself to Gibbs sampling.  We sample the posterior probability for the signal and signal covariance $P(s,\mathbf{S}|d)$ by iteratively sampling the conditional distributions
\begin{eqnarray}
  \label{eqn:gibbs}
  t &\leftarrow& P(t|d,s,\mathbf{S})  \\
  s &\leftarrow& P(s|d,t,\mathbf{S})  \nonumber \\
  \mathbf{S} &\leftarrow& P(\mathbf{S}|d,s,t) = P(\mathbf{S}|s).  \nonumber
\end{eqnarray}
The first two conditional distributions are Gaussian and arise from completing the square, and the third is an inverse Gamma (or inverse Wishart) distribution \citep{2007ApJ...656..653L}.
  
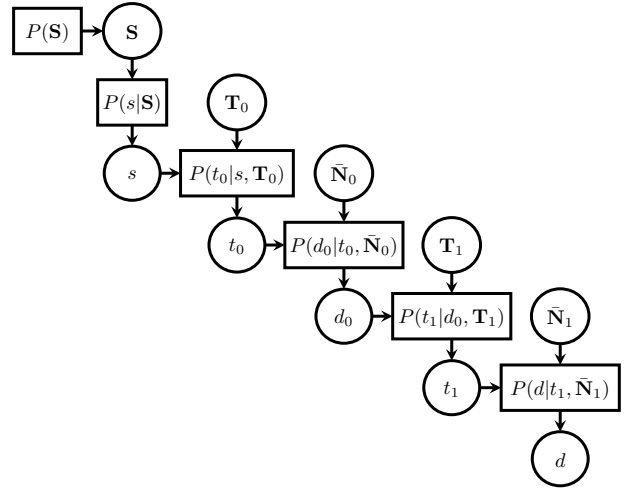
\begin{figure}
  \begin{center}
    \resizebox{\columnwidth}{!}{
      \large
      \input{figs/forward_model_clean.tex}
    }
  \end{center}
  \caption{Hierarchical forward model for multiple messenger fields, starting with the prior on the signal covariance, and proceeding from the signal ($s$) via the messenger and auxiliary fields ($t_0,d_0,t_1$) to the final data ($d$).  This shows how we accumulate noise such that $d_0$ has a noise with covariance $\mathbf{N}_0 = \bar \mathbf{N}_0 + \mathbf{T}_0$ and $d$ has noise with covariance $\mathbf{N} = \mathbf{N}_0 + \mathbf{N}_1$.} \label{fig:hierarchy}
\end{figure}

The original EW work considered the total covariance as the sum of two parts, for the signal and the noise.  The equations they derive are always valid, but we have noted that the method is really suited to the case when the noise and signal covariance are each sparse in convenient bases.  Here we generalize to 
show that if one of the covariances is still dense, but decomposes into pieces that are sparse, we can continue to break it up until we get pieces that are easily invertible.



We proceed by adding additional fields.  We treat the case with  $\mathbf{N} = \mathbf{N}_0 + \mathbf{N}_1$, but the idea generalizes easily if more parts are needed to represent the covariance.  A hierarchical forward model in this case (Fig.~\ref{fig:hierarchy}) builds up the noise bit-by-bit.  The Gaussian probability corresponding to this diagram is
\begin{eqnarray}
  \label{eqn:2mess_prob}
  -2 \log P_2(s,t_0,d_0,t_1,d) &=& s^\dag \mathbf{S}^{-1} s \\ \nonumber
  &&+ (t_0-s)^\dag\mathbf{T}_0^{-1}(t_0-s) \\ \nonumber
  &&+ (d_0-t_0)^\dag \mathbf{\bar N}_0^{-1}(d_0-t_0) \\ \nonumber
  &&+ (t_1-d_0)^\dag\mathbf{T}_1^{-1}(t_1-d_0) \\ \nonumber
  &&+ (d-t_1)^\dag \mathbf{\bar N}_1^{-1}(d-t_1) \\ \nonumber
  &&+ \mbox{const.}
\end{eqnarray}
where $ \bar \mathbf{N}_0 = \mathbf{N}_0 - \mathbf{T}_0$ and $ \bar \mathbf{N}_1= \mathbf{N}_1 - \mathbf{T}_1$.

In the same way as before, we complete the square for the signal, messenger, and auxiliary fields in (\ref{eqn:2mess_prob}).  This provides in turn the Gaussian probability distributions for each variable conditioned on the others.  The means of these distributions yield our iterative solution to the Wiener filter, equations~(\ref{eqn:2mess_t1})--(\ref{eqn:2mess_s}).  The full distributions can be used for Gibbs sampling, in analogy to equation~(\ref{eqn:gibbs}).

As with EW, smart choices for the covariances of the messenger fields are proportional to the identity matrix so that they are sparse in any basis,
\begin{eqnarray}
  \mathbf{T_0} &=& \tau_0 \mathbf{I} \\ \nonumber
  \mathbf{T_1} &=& \tau_1 \mathbf{I}.
\end{eqnarray}
Thus the messenger field $t_0$ represents $s$ plus some uniform noise; auxiliary field $d_0$ includes the remainder of the $\mathbf{N}_0$ noise; messenger field $t_1$ includes some more uniform noise; and $d$ includes the remainder of the $\mathbf{N}_1$ noise.

Because we compute both equations~(\ref{eqn:2mess_d0})--(\ref{eqn:2mess_t0}) in the sparse basis for $\mathbf{N}_0$, they can be combined as
\begin{eqnarray}
  t_0 = [ \bar  \mathbf{N}_0^{-1} +  \mathbf{T}_0^{-1} -\mathbf{N}_0^{-1}(\bar  \mathbf{N}_0^{-1} +  \mathbf{T}_1^{-1}) \mathbf{N}_0^{-1} ]^{-1} \\ \nonumber \left( \mathbf{ \bar N}_0^{-1}(\bar  \mathbf{N}_0^{-1} +  \mathbf{T}_0^{-1})^{-1}  \mathbf{ T}_1^{-1} t_1 + \mathbf{ T}_0^{-1} s \right),
\end{eqnarray}
which solves for $t_0$ in terms of $t_1$ and $s$.  Note also that equations~(\ref{eqn:2mess_t0})--(\ref{eqn:2mess_s}) resemble the EW solution for the Wiener filter, and so 
\begin{equation}
  s = (\mathbf{S}^{-1} + \mathbf{N}_0^{-1})^{-1} \mathbf{ N}_0^{-1} d_0,
\end{equation}
useful in the specific case where the signal covariance is sparse in the same basis as the $\mathbf{N}_0$ portion of the noise.  These shortcuts may speed up the convergence in some cases.

\bibliographystyle{mnras}
\bibliography{ref}

\end{document}

%% file: figs/forward_model_clean.tex
\ifx\du\undefined
  \newlength{\du}
\fi
\setlength{\du}{15\unitlength}
\begin{tikzpicture}
\pgftransformxscale{1.000000}
\pgftransformyscale{-1.000000}
\definecolor{dialinecolor}{rgb}{0.000000, 0.000000, 0.000000}
\pgfsetstrokecolor{dialinecolor}
\definecolor{dialinecolor}{rgb}{1.000000, 1.000000, 1.000000}
\pgfsetfillcolor{dialinecolor}
\definecolor{dialinecolor}{rgb}{1.000000, 1.000000, 1.000000}
\pgfsetfillcolor{dialinecolor}
\fill (-0.000160\du,0.697815\du)--(-0.000160\du,2.221148\du)--(2.252340\du,2.221148\du)--(2.252340\du,0.697815\du)--cycle;
\pgfsetlinewidth{0.100000\du}
\pgfsetdash{}{0pt}
\pgfsetdash{}{0pt}
\pgfsetmiterjoin
\definecolor{dialinecolor}{rgb}{0.000000, 0.000000, 0.000000}
\pgfsetstrokecolor{dialinecolor}
\draw (-0.000160\du,0.697815\du)--(-0.000160\du,2.221148\du)--(2.252340\du,2.221148\du)--(2.252340\du,0.697815\du)--cycle;
\definecolor{dialinecolor}{rgb}{0.000000, 0.000000, 0.000000}
\pgfsetstrokecolor{dialinecolor}
\node at (1.126090\du,1.511148\du){$P(\mathbf{S})$};
\definecolor{dialinecolor}{rgb}{1.000000, 1.000000, 1.000000}
\pgfsetfillcolor{dialinecolor}
\pgfpathellipse{\pgfpoint{3.944828\du}{6.228038\du}}{\pgfpoint{0.929752\du}{0\du}}{\pgfpoint{0\du}{0.906479\du}}
\pgfusepath{fill}
\pgfsetlinewidth{0.100000\du}
\pgfsetdash{}{0pt}
\pgfsetdash{}{0pt}
\pgfsetmiterjoin
\definecolor{dialinecolor}{rgb}{0.000000, 0.000000, 0.000000}
\pgfsetstrokecolor{dialinecolor}
\pgfpathellipse{\pgfpoint{3.944828\du}{6.228038\du}}{\pgfpoint{0.929752\du}{0\du}}{\pgfpoint{0\du}{0.906479\du}}
\pgfusepath{stroke}
\definecolor{dialinecolor}{rgb}{0.000000, 0.000000, 0.000000}
\pgfsetstrokecolor{dialinecolor}
\node at (3.944828\du,6.279704\du){$s$};
\definecolor{dialinecolor}{rgb}{1.000000, 1.000000, 1.000000}
\pgfsetfillcolor{dialinecolor}
\fill (2.796564\du,3.092107\du)--(2.796564\du,4.615440\du)--(5.093091\du,4.615440\du)--(5.093091\du,3.092107\du)--cycle;
\pgfsetlinewidth{0.100000\du}
\pgfsetdash{}{0pt}
\pgfsetdash{}{0pt}
\pgfsetmiterjoin
\definecolor{dialinecolor}{rgb}{0.000000, 0.000000, 0.000000}
\pgfsetstrokecolor{dialinecolor}
\draw (2.796564\du,3.092107\du)--(2.796564\du,4.615440\du)--(5.093091\du,4.615440\du)--(5.093091\du,3.092107\du)--cycle;
\definecolor{dialinecolor}{rgb}{0.000000, 0.000000, 0.000000}
\pgfsetstrokecolor{dialinecolor}
\node at (3.944828\du,3.905440\du){$P(s|\mathbf{S})$};
\definecolor{dialinecolor}{rgb}{1.000000, 1.000000, 1.000000}
\pgfsetfillcolor{dialinecolor}
\pgfpathellipse{\pgfpoint{3.944828\du}{1.459481\du}}{\pgfpoint{0.957886\du}{0\du}}{\pgfpoint{0\du}{0.919224\du}}
\pgfusepath{fill}
\pgfsetlinewidth{0.100000\du}
\pgfsetdash{}{0pt}
\pgfsetdash{}{0pt}
\pgfsetmiterjoin
\definecolor{dialinecolor}{rgb}{0.000000, 0.000000, 0.000000}
\pgfsetstrokecolor{dialinecolor}
\pgfpathellipse{\pgfpoint{3.944828\du}{1.459481\du}}{\pgfpoint{0.957886\du}{0\du}}{\pgfpoint{0\du}{0.919224\du}}
\pgfusepath{stroke}
\definecolor{dialinecolor}{rgb}{0.000000, 0.000000, 0.000000}
\pgfsetstrokecolor{dialinecolor}
\node at (3.944828\du,1.511148\du){$\mathbf{S}$};
\pgfsetlinewidth{0.100000\du}
\pgfsetdash{}{0pt}
\pgfsetdash{}{0pt}
\pgfsetbuttcap
{
\definecolor{dialinecolor}{rgb}{0.000000, 0.000000, 0.000000}
\pgfsetfillcolor{dialinecolor}
\pgfsetarrowsend{stealth}
\definecolor{dialinecolor}{rgb}{0.000000, 0.000000, 0.000000}
\pgfsetstrokecolor{dialinecolor}
\draw (3.944828\du,2.378706\du)--(3.944828\du,3.092107\du);
}
\definecolor{dialinecolor}{rgb}{1.000000, 1.000000, 1.000000}
\pgfsetfillcolor{dialinecolor}
\fill (5.553811\du,5.466371\du)--(5.553811\du,6.989704\du)--(9.306402\du,6.989704\du)--(9.306402\du,5.466371\du)--cycle;
\pgfsetlinewidth{0.100000\du}
\pgfsetdash{}{0pt}
\pgfsetdash{}{0pt}
\pgfsetmiterjoin
\definecolor{dialinecolor}{rgb}{0.000000, 0.000000, 0.000000}
\pgfsetstrokecolor{dialinecolor}
\draw (5.553811\du,5.466371\du)--(5.553811\du,6.989704\du)--(9.306402\du,6.989704\du)--(9.306402\du,5.466371\du)--cycle;
\definecolor{dialinecolor}{rgb}{0.000000, 0.000000, 0.000000}
\pgfsetstrokecolor{dialinecolor}
\node at (7.430106\du,6.279704\du){$P(t_0|s,\mathbf{T}_0)$};
\pgfsetlinewidth{0.100000\du}
\pgfsetdash{}{0pt}
\pgfsetdash{}{0pt}
\pgfsetbuttcap
{
\definecolor{dialinecolor}{rgb}{0.000000, 0.000000, 0.000000}
\pgfsetfillcolor{dialinecolor}
\pgfsetarrowsend{stealth}
\definecolor{dialinecolor}{rgb}{0.000000, 0.000000, 0.000000}
\pgfsetstrokecolor{dialinecolor}
\draw (4.874580\du,6.228038\du)--(5.553811\du,6.228038\du);
}
\definecolor{dialinecolor}{rgb}{1.000000, 1.000000, 1.000000}
\pgfsetfillcolor{dialinecolor}
\pgfpathellipse{\pgfpoint{7.430106\du}{3.853773\du}}{\pgfpoint{0.957886\du}{0\du}}{\pgfpoint{0\du}{0.919224\du}}
\pgfusepath{fill}
\pgfsetlinewidth{0.100000\du}
\pgfsetdash{}{0pt}
\pgfsetdash{}{0pt}
\pgfsetmiterjoin
\definecolor{dialinecolor}{rgb}{0.000000, 0.000000, 0.000000}
\pgfsetstrokecolor{dialinecolor}
\pgfpathellipse{\pgfpoint{7.430106\du}{3.853773\du}}{\pgfpoint{0.957886\du}{0\du}}{\pgfpoint{0\du}{0.919224\du}}
\pgfusepath{stroke}
\definecolor{dialinecolor}{rgb}{0.000000, 0.000000, 0.000000}
\pgfsetstrokecolor{dialinecolor}
\node at (7.430106\du,3.905440\du){$\mathbf{T}_0$};
\pgfsetlinewidth{0.100000\du}
\pgfsetdash{}{0pt}
\pgfsetdash{}{0pt}
\pgfsetbuttcap
{
\definecolor{dialinecolor}{rgb}{0.000000, 0.000000, 0.000000}
\pgfsetfillcolor{dialinecolor}
\pgfsetarrowsend{stealth}
\definecolor{dialinecolor}{rgb}{0.000000, 0.000000, 0.000000}
\pgfsetstrokecolor{dialinecolor}
\draw (7.430106\du,4.772998\du)--(7.430106\du,5.466371\du);
}
\definecolor{dialinecolor}{rgb}{1.000000, 1.000000, 1.000000}
\pgfsetfillcolor{dialinecolor}
\pgfpathellipse{\pgfpoint{7.430106\du}{8.634589\du}}{\pgfpoint{0.929752\du}{0\du}}{\pgfpoint{0\du}{0.906479\du}}
\pgfusepath{fill}
\pgfsetlinewidth{0.100000\du}
\pgfsetdash{}{0pt}
\pgfsetdash{}{0pt}
\pgfsetmiterjoin
\definecolor{dialinecolor}{rgb}{0.000000, 0.000000, 0.000000}
\pgfsetstrokecolor{dialinecolor}
\pgfpathellipse{\pgfpoint{7.430106\du}{8.634589\du}}{\pgfpoint{0.929752\du}{0\du}}{\pgfpoint{0\du}{0.906479\du}}
\pgfusepath{stroke}
\definecolor{dialinecolor}{rgb}{0.000000, 0.000000, 0.000000}
\pgfsetstrokecolor{dialinecolor}
\node at (7.430106\du,8.686256\du){$t_0$};
\pgfsetlinewidth{0.100000\du}
\pgfsetdash{}{0pt}
\pgfsetdash{}{0pt}
\pgfsetbuttcap
{
\definecolor{dialinecolor}{rgb}{0.000000, 0.000000, 0.000000}
\pgfsetfillcolor{dialinecolor}
\pgfsetarrowsend{stealth}
\definecolor{dialinecolor}{rgb}{0.000000, 0.000000, 0.000000}
\pgfsetstrokecolor{dialinecolor}
\draw (7.430106\du,6.989704\du)--(7.430106\du,7.728110\du);
}
\definecolor{dialinecolor}{rgb}{1.000000, 1.000000, 1.000000}
\pgfsetfillcolor{dialinecolor}
\pgfpathellipse{\pgfpoint{14.576527\du}{8.634589\du}}{\pgfpoint{0.957886\du}{0\du}}{\pgfpoint{0\du}{0.919224\du}}
\pgfusepath{fill}
\pgfsetlinewidth{0.100000\du}
\pgfsetdash{}{0pt}
\pgfsetdash{}{0pt}
\pgfsetmiterjoin
\definecolor{dialinecolor}{rgb}{0.000000, 0.000000, 0.000000}
\pgfsetstrokecolor{dialinecolor}
\pgfpathellipse{\pgfpoint{14.576527\du}{8.634589\du}}{\pgfpoint{0.957886\du}{0\du}}{\pgfpoint{0\du}{0.919224\du}}
\pgfusepath{stroke}
\definecolor{dialinecolor}{rgb}{0.000000, 0.000000, 0.000000}
\pgfsetstrokecolor{dialinecolor}
\node at (14.576527\du,8.686256\du){$\mathbf{T}_1$};
\definecolor{dialinecolor}{rgb}{1.000000, 1.000000, 1.000000}
\pgfsetfillcolor{dialinecolor}
\pgfpathellipse{\pgfpoint{18.185551\du}{11.021397\du}}{\pgfpoint{0.957886\du}{0\du}}{\pgfpoint{0\du}{0.919224\du}}
\pgfusepath{fill}
\pgfsetlinewidth{0.100000\du}
\pgfsetdash{}{0pt}
\pgfsetdash{}{0pt}
\pgfsetmiterjoin
\definecolor{dialinecolor}{rgb}{0.000000, 0.000000, 0.000000}
\pgfsetstrokecolor{dialinecolor}
\pgfpathellipse{\pgfpoint{18.185551\du}{11.021397\du}}{\pgfpoint{0.957886\du}{0\du}}{\pgfpoint{0\du}{0.919224\du}}
\pgfusepath{stroke}
\definecolor{dialinecolor}{rgb}{0.000000, 0.000000, 0.000000}
\pgfsetstrokecolor{dialinecolor}
\node at (18.185551\du,11.073063\du){$\bar \mathbf{N}_1$};
\definecolor{dialinecolor}{rgb}{1.000000, 1.000000, 1.000000}
\pgfsetfillcolor{dialinecolor}
\pgfpathellipse{\pgfpoint{10.988747\du}{6.228038\du}}{\pgfpoint{0.957886\du}{0\du}}{\pgfpoint{0\du}{0.919224\du}}
\pgfusepath{fill}
\pgfsetlinewidth{0.100000\du}
\pgfsetdash{}{0pt}
\pgfsetdash{}{0pt}
\pgfsetmiterjoin
\definecolor{dialinecolor}{rgb}{0.000000, 0.000000, 0.000000}
\pgfsetstrokecolor{dialinecolor}
\pgfpathellipse{\pgfpoint{10.988747\du}{6.228038\du}}{\pgfpoint{0.957886\du}{0\du}}{\pgfpoint{0\du}{0.919224\du}}
\pgfusepath{stroke}
\definecolor{dialinecolor}{rgb}{0.000000, 0.000000, 0.000000}
\pgfsetstrokecolor{dialinecolor}
\node at (10.988747\du,6.279704\du){$\bar \mathbf{N}_0$};
\definecolor{dialinecolor}{rgb}{1.000000, 1.000000, 1.000000}
\pgfsetfillcolor{dialinecolor}
\fill (9.023284\du,7.872922\du)--(9.023284\du,9.396256\du)--(12.954209\du,9.396256\du)--(12.954209\du,7.872922\du)--cycle;
\pgfsetlinewidth{0.100000\du}
\pgfsetdash{}{0pt}
\pgfsetdash{}{0pt}
\pgfsetmiterjoin
\definecolor{dialinecolor}{rgb}{0.000000, 0.000000, 0.000000}
\pgfsetstrokecolor{dialinecolor}
\draw (9.023284\du,7.872922\du)--(9.023284\du,9.396256\du)--(12.954209\du,9.396256\du)--(12.954209\du,7.872922\du)--cycle;
\definecolor{dialinecolor}{rgb}{0.000000, 0.000000, 0.000000}
\pgfsetstrokecolor{dialinecolor}
\node at (10.988747\du,8.686256\du){$P(d_0|t_0,\bar \mathbf{N}_0)$};
\definecolor{dialinecolor}{rgb}{1.000000, 1.000000, 1.000000}
\pgfsetfillcolor{dialinecolor}
\fill (12.611065\du,10.259730\du)--(12.611065\du,11.783063\du)--(16.541990\du,11.783063\du)--(16.541990\du,10.259730\du)--cycle;
\pgfsetlinewidth{0.100000\du}
\pgfsetdash{}{0pt}
\pgfsetdash{}{0pt}
\pgfsetmiterjoin
\definecolor{dialinecolor}{rgb}{0.000000, 0.000000, 0.000000}
\pgfsetstrokecolor{dialinecolor}
\draw (12.611065\du,10.259730\du)--(12.611065\du,11.783063\du)--(16.541990\du,11.783063\du)--(16.541990\du,10.259730\du)--cycle;
\definecolor{dialinecolor}{rgb}{0.000000, 0.000000, 0.000000}
\pgfsetstrokecolor{dialinecolor}
\node at (14.576527\du,11.073063\du){$P(t_1|d_0,\mathbf{T}_1)$};
\definecolor{dialinecolor}{rgb}{1.000000, 1.000000, 1.000000}
\pgfsetfillcolor{dialinecolor}
\fill (16.220089\du,12.655914\du)--(16.220089\du,14.179247\du)--(20.151014\du,14.179247\du)--(20.151014\du,12.655914\du)--cycle;
\pgfsetlinewidth{0.100000\du}
\pgfsetdash{}{0pt}
\pgfsetdash{}{0pt}
\pgfsetmiterjoin
\definecolor{dialinecolor}{rgb}{0.000000, 0.000000, 0.000000}
\pgfsetstrokecolor{dialinecolor}
\draw (16.220089\du,12.655914\du)--(16.220089\du,14.179247\du)--(20.151014\du,14.179247\du)--(20.151014\du,12.655914\du)--cycle;
\definecolor{dialinecolor}{rgb}{0.000000, 0.000000, 0.000000}
\pgfsetstrokecolor{dialinecolor}
\node at (18.185551\du,13.469247\du){$P(d|t_1,\bar \mathbf{N}_1)$};
\definecolor{dialinecolor}{rgb}{1.000000, 1.000000, 1.000000}
\pgfsetfillcolor{dialinecolor}
\pgfpathellipse{\pgfpoint{14.576527\du}{13.417581\du}}{\pgfpoint{0.929752\du}{0\du}}{\pgfpoint{0\du}{0.906479\du}}
\pgfusepath{fill}
\pgfsetlinewidth{0.100000\du}
\pgfsetdash{}{0pt}
\pgfsetdash{}{0pt}
\pgfsetmiterjoin
\definecolor{dialinecolor}{rgb}{0.000000, 0.000000, 0.000000}
\pgfsetstrokecolor{dialinecolor}
\pgfpathellipse{\pgfpoint{14.576527\du}{13.417581\du}}{\pgfpoint{0.929752\du}{0\du}}{\pgfpoint{0\du}{0.906479\du}}
\pgfusepath{stroke}
\definecolor{dialinecolor}{rgb}{0.000000, 0.000000, 0.000000}
\pgfsetstrokecolor{dialinecolor}
\node at (14.576527\du,13.469247\du){$t_1$};
\definecolor{dialinecolor}{rgb}{1.000000, 1.000000, 1.000000}
\pgfsetfillcolor{dialinecolor}
\pgfpathellipse{\pgfpoint{18.185551\du}{15.800101\du}}{\pgfpoint{0.929752\du}{0\du}}{\pgfpoint{0\du}{0.906479\du}}
\pgfusepath{fill}
\pgfsetlinewidth{0.100000\du}
\pgfsetdash{}{0pt}
\pgfsetdash{}{0pt}
\pgfsetmiterjoin
\definecolor{dialinecolor}{rgb}{0.000000, 0.000000, 0.000000}
\pgfsetstrokecolor{dialinecolor}
\pgfpathellipse{\pgfpoint{18.185551\du}{15.800101\du}}{\pgfpoint{0.929752\du}{0\du}}{\pgfpoint{0\du}{0.906479\du}}
\pgfusepath{stroke}
\definecolor{dialinecolor}{rgb}{0.000000, 0.000000, 0.000000}
\pgfsetstrokecolor{dialinecolor}
\node at (18.185551\du,15.851767\du){$d$};
\pgfsetlinewidth{0.100000\du}
\pgfsetdash{}{0pt}
\pgfsetdash{}{0pt}
\pgfsetbuttcap
{
\definecolor{dialinecolor}{rgb}{0.000000, 0.000000, 0.000000}
\pgfsetfillcolor{dialinecolor}
\pgfsetarrowsend{stealth}
\definecolor{dialinecolor}{rgb}{0.000000, 0.000000, 0.000000}
\pgfsetstrokecolor{dialinecolor}
\draw (18.185551\du,14.179247\du)--(18.185551\du,14.893621\du);
}
\pgfsetlinewidth{0.100000\du}
\pgfsetdash{}{0pt}
\pgfsetdash{}{0pt}
\pgfsetbuttcap
{
\definecolor{dialinecolor}{rgb}{0.000000, 0.000000, 0.000000}
\pgfsetfillcolor{dialinecolor}
\pgfsetarrowsend{stealth}
\definecolor{dialinecolor}{rgb}{0.000000, 0.000000, 0.000000}
\pgfsetstrokecolor{dialinecolor}
\draw (14.576527\du,11.783063\du)--(14.576527\du,12.511101\du);
}
\pgfsetlinewidth{0.100000\du}
\pgfsetdash{}{0pt}
\pgfsetdash{}{0pt}
\pgfsetbuttcap
{
\definecolor{dialinecolor}{rgb}{0.000000, 0.000000, 0.000000}
\pgfsetfillcolor{dialinecolor}
\pgfsetarrowsend{stealth}
\definecolor{dialinecolor}{rgb}{0.000000, 0.000000, 0.000000}
\pgfsetstrokecolor{dialinecolor}
\draw (15.506280\du,13.417581\du)--(16.220089\du,13.417581\du);
}
\pgfsetlinewidth{0.100000\du}
\pgfsetdash{}{0pt}
\pgfsetdash{}{0pt}
\pgfsetbuttcap
{
\definecolor{dialinecolor}{rgb}{0.000000, 0.000000, 0.000000}
\pgfsetfillcolor{dialinecolor}
\pgfsetarrowsend{stealth}
\definecolor{dialinecolor}{rgb}{0.000000, 0.000000, 0.000000}
\pgfsetstrokecolor{dialinecolor}
\draw (18.185551\du,11.940621\du)--(18.185551\du,12.655914\du);
}
\pgfsetlinewidth{0.100000\du}
\pgfsetdash{}{0pt}
\pgfsetdash{}{0pt}
\pgfsetbuttcap
{
\definecolor{dialinecolor}{rgb}{0.000000, 0.000000, 0.000000}
\pgfsetfillcolor{dialinecolor}
\pgfsetarrowsend{stealth}
\definecolor{dialinecolor}{rgb}{0.000000, 0.000000, 0.000000}
\pgfsetstrokecolor{dialinecolor}
\draw (10.988747\du,7.147262\du)--(10.988747\du,7.872922\du);
}
\pgfsetlinewidth{0.100000\du}
\pgfsetdash{}{0pt}
\pgfsetdash{}{0pt}
\pgfsetbuttcap
{
\definecolor{dialinecolor}{rgb}{0.000000, 0.000000, 0.000000}
\pgfsetfillcolor{dialinecolor}
\pgfsetarrowsend{stealth}
\definecolor{dialinecolor}{rgb}{0.000000, 0.000000, 0.000000}
\pgfsetstrokecolor{dialinecolor}
\draw (14.576527\du,9.553814\du)--(14.576527\du,10.259730\du);
}
\pgfsetlinewidth{0.100000\du}
\pgfsetdash{}{0pt}
\pgfsetdash{}{0pt}
\pgfsetbuttcap
{
\definecolor{dialinecolor}{rgb}{0.000000, 0.000000, 0.000000}
\pgfsetfillcolor{dialinecolor}
\pgfsetarrowsend{stealth}
\definecolor{dialinecolor}{rgb}{0.000000, 0.000000, 0.000000}
\pgfsetstrokecolor{dialinecolor}
\draw (8.359859\du,8.634589\du)--(9.023284\du,8.634589\du);
}
\pgfsetlinewidth{0.100000\du}
\pgfsetdash{}{0pt}
\pgfsetdash{}{0pt}
\pgfsetbuttcap
{
\definecolor{dialinecolor}{rgb}{0.000000, 0.000000, 0.000000}
\pgfsetfillcolor{dialinecolor}
\pgfsetarrowsend{stealth}
\definecolor{dialinecolor}{rgb}{0.000000, 0.000000, 0.000000}
\pgfsetstrokecolor{dialinecolor}
\draw (3.944828\du,4.615440\du)--(3.944828\du,5.321559\du);
}
\pgfsetlinewidth{0.100000\du}
\pgfsetdash{}{0pt}
\pgfsetdash{}{0pt}
\pgfsetbuttcap
{
\definecolor{dialinecolor}{rgb}{0.000000, 0.000000, 0.000000}
\pgfsetfillcolor{dialinecolor}
\pgfsetarrowsend{stealth}
\definecolor{dialinecolor}{rgb}{0.000000, 0.000000, 0.000000}
\pgfsetstrokecolor{dialinecolor}
\draw (2.252340\du,1.459481\du)--(2.986942\du,1.459481\du);
}
\definecolor{dialinecolor}{rgb}{1.000000, 1.000000, 1.000000}
\pgfsetfillcolor{dialinecolor}
\pgfpathellipse{\pgfpoint{10.988747\du}{11.021397\du}}{\pgfpoint{0.929752\du}{0\du}}{\pgfpoint{0\du}{0.906479\du}}
\pgfusepath{fill}
\pgfsetlinewidth{0.100000\du}
\pgfsetdash{}{0pt}
\pgfsetdash{}{0pt}
\pgfsetmiterjoin
\definecolor{dialinecolor}{rgb}{0.000000, 0.000000, 0.000000}
\pgfsetstrokecolor{dialinecolor}
\pgfpathellipse{\pgfpoint{10.988747\du}{11.021397\du}}{\pgfpoint{0.929752\du}{0\du}}{\pgfpoint{0\du}{0.906479\du}}
\pgfusepath{stroke}
\definecolor{dialinecolor}{rgb}{0.000000, 0.000000, 0.000000}
\pgfsetstrokecolor{dialinecolor}
\node at (10.988747\du,11.073063\du){$d_0$};
\pgfsetlinewidth{0.100000\du}
\pgfsetdash{}{0pt}
\pgfsetdash{}{0pt}
\pgfsetbuttcap
{
\definecolor{dialinecolor}{rgb}{0.000000, 0.000000, 0.000000}
\pgfsetfillcolor{dialinecolor}
\pgfsetarrowsend{stealth}
\definecolor{dialinecolor}{rgb}{0.000000, 0.000000, 0.000000}
\pgfsetstrokecolor{dialinecolor}
\draw (10.988747\du,9.396256\du)--(10.988747\du,10.114918\du);
}
\pgfsetlinewidth{0.100000\du}
\pgfsetdash{}{0pt}
\pgfsetdash{}{0pt}
\pgfsetbuttcap
{
\definecolor{dialinecolor}{rgb}{0.000000, 0.000000, 0.000000}
\pgfsetfillcolor{dialinecolor}
\pgfsetarrowsend{stealth}
\definecolor{dialinecolor}{rgb}{0.000000, 0.000000, 0.000000}
\pgfsetstrokecolor{dialinecolor}
\draw (11.918499\du,11.021397\du)--(12.611065\du,11.021397\du);
}
\end{tikzpicture}